\documentclass[journal]{IEEEtranTIE}
\usepackage{graphicx}
\usepackage{cite}
\usepackage{picinpar}
\usepackage{amsmath}
\usepackage{url}
\usepackage{flushend}
\usepackage{colortbl}
\usepackage{soul}
\usepackage{multirow}
\usepackage{pifont}
\usepackage{color}
\usepackage[table,xcdraw]{xcolor}
\usepackage{xcolor}
\usepackage{alltt}
\usepackage[hidelinks]{hyperref}
\usepackage{enumerate}
\usepackage{siunitx}
\usepackage{breakurl}
\usepackage{epstopdf}
\usepackage{pbox}
\usepackage{amsmath,amsfonts}
\usepackage{algorithmic}
\usepackage{algorithm}
\usepackage{array}
\usepackage[caption=false,font=normalsize,labelfont=sf,textfont=sf]{subfig}
\usepackage{textcomp}
\usepackage{stfloats}
\usepackage{url}
\usepackage{verbatim}
\usepackage{graphicx}
\usepackage{cite}
\usepackage{breqn}
\usepackage{makecell}
\usepackage[utf8]{inputenc} % UTF-8 인코딩 사용

\begin{document}
\title{	Temperature Compensation Method of Six-Axis Force/Torque Sensor Using Gated Recurrent Unit \\ (Feb. 2025)}

\author{
	\vskip 1em
	
	Hyun-Bin Kim, Seokju Lee, Byeong-Il Ham, and Kyung-Soo Kim,~\IEEEmembership{Member,~IEEE,}

	\thanks{
	 Manuscript created September, 2024; This work was developed by the MSC (Mechatronics, Systems and Control) lab in the KAIST(Korea Advanced Institute of Science and Technology which is in the Daehak-Ro 291, Daejeon, South Korea(e-mail: youfree22@kaist.ac.kr; dltjrwn0322@kaist.ac.kr; byeongil\_ham@kaist.ac.kr; kyungsookim@kaist.ac.kr).(Corresponding author: Kyung-Soo Kim).  
	}
}

\maketitle
	
\begin{abstract}
This study aims to enhance the accuracy of a six-axis force/torque sensor compared to existing approaches that utilize Multi-Layer Perceptron (MLP) and the Least Square Method. The sensor used in this study is based on a photo-coupler and operates with infrared light, making it susceptible to dark current effects, which cause drift due to temperature variations. Additionally, the sensor is compact and lightweight (45g), resulting in a low thermal capacity. Consequently, even small amounts of heat can induce rapid temperature changes, affecting the sensor's performance in real time. To address these challenges, this study compares the conventional MLP approach with the proposed Gated Recurrent Unit (GRU)-based method. Experimental results demonstrate that the GRU approach, leveraging sequential data, achieves superior performance.
\end{abstract}


\markboth{arXiv}%
{}

\definecolor{limegreen}{rgb}{0.2, 0.8, 0.2}
\definecolor{forestgreen}{rgb}{0.13, 0.55, 0.13}
\definecolor{greenhtml}{rgb}{0.0, 0.5, 0.0}

\section{Introduction}

\IEEEPARstart{R}{ecently}, various types of robots, including collaborative robots, legged robots, and mobile robots, have been actively researched~\cite{di2018dynamic,hutter2016anymal,kang2023external,kang2024external,kang2023view}. Force/Torque (F/T) sensors are widely used in these robots~\cite{li2025impedance,valsecchi2020quadrupedal}, with strain gauge-based sensors being the most common. In addition to strain gauges, capacitance-based F/T sensors are also utilized~\cite{kim2016novel,kim2018sensorized,pu2021modeling}, as they enable precise measurement of forces and torques.

These sensors serve various purposes in different robotic applications. In collaborative robots, they are employed for admittance control~\cite{chen2025adaptive}, grasping objects, and determining the position of the end-effector. In legged robots, particularly bipedal robots, F/T sensors are frequently used to measure ground reaction forces, which play a crucial role in walking control by enabling the calculation of Zero Moment Point (ZMP)~\cite{kajita2003biped} or Capture Point~\cite{pratt2006capture}. In mobile robots, these sensors are often attached to robotic arms for object manipulation. In quadrupedal robots, pneumatic pressure sensors are commonly used to detect contact~\cite{unitree_website}.

However, F/T sensors, particularly those based on strain gauges, are susceptible to drift or offset due to thermal expansion caused by temperature variations. Various methods have been proposed to address this issue, which can be broadly categorized into two approaches. The first is model-based methods, while the second involves the use of artificial neural networks or support vector machines.

Model-based methods primarily utilize linear approaches, incorporating a calibration matrix and a function of temperature as follows:

\begin{equation}
f=\textbf{C}r+o+\textbf{C}_t t    
\end{equation}
~\cite{scott2011novel}
where \(\textbf{C}\) represents the calibration matrix, \(o\) denotes the offset, \(\textbf{C}_t\) refers to the temperature calibration coefficients, and \(t\) is the temperature value. This method employs the Least Square Method, and similar approaches have been used in \cite{chavez2019model,scott2011novel,billeschou2021low}. In fiber grating-based methods, an analytical model-based approach has been implemented using heat transfer models~\cite{wang2021improved} or the Least Square Method. Additionally, for piezoresistive sensors~\cite{wang2017temperature}, model-based methods incorporating the characteristics of the Wheatstone bridge have been applied. Linear models have also been utilized in some studies.

Research employing Support Vector Machines (SVM) and Artificial Neural Networks (ANN)~\cite{kazemi2020temperature} for thermal compensation includes various approaches. One such approach involves using a Multi-Layer Perceptron (MLP) with Radial Basis Functions (RBF)~\cite{sun2014temperature}. Subsequent studies have optimized RBF networks using particle swarm optimization and applied the Least Square Support Vector Machine (LSSVM)~\cite{sun2015temperature}. Other approaches include employing MLP with Leaky ReLU activation functions or optimizing using the Adaptive Genetic Algorithm-Back Propagation (AGA-BP) method~\cite{han2020temperature}. Moreover, additional studies have implemented thermal compensation using SVM, K-nearest neighbors (KNN), Random Forest, and Decision Trees~\cite{kazemi2021comparative}.

However, in most of these studies, modeling is often simplified, assuming no axis interference or that the sensor weight is in the kilogram range, allowing thermal compensation to reach a steady state. Linear models introduce significant compensation errors when approximating nonlinear drift as a linear model. Furthermore, for heavy sensors that maintain a steady-state temperature, the use of MLP does not lead to significant errors.

This study aims to address temperature-induced drift in a small sensor (diameter 40 mm, 45 g weight)~\cite{kim2025parameter,kim2024compact} that experiences rapid temperature variations due to low thermal mass. The key contributions of this work are as follows. To compensate for sensor drift, time sequence data is utilized for training, and a Gated Recurrent Unit (GRU)~\cite{cho2014learning}, a type of Recurrent Neural Network (RNN) commonly used in natural language models, is employed. Experiments compare the proposed approach with the Least Square Method and MLP-based training, demonstrating that the proposed method exhibits lower errors and superior performance in scenarios where temperature continuously varies over time.

This approach has the potential to be applied to various sensors, particularly in cases where nonlinearities arise due to temperature variations.
\section{A NOVEL SIX-AXIS FORCE/TORQUE SENSOR USING PHOTO-COUPLERs}
\begin{figure}
    \centering
    \includegraphics[width=1\linewidth]{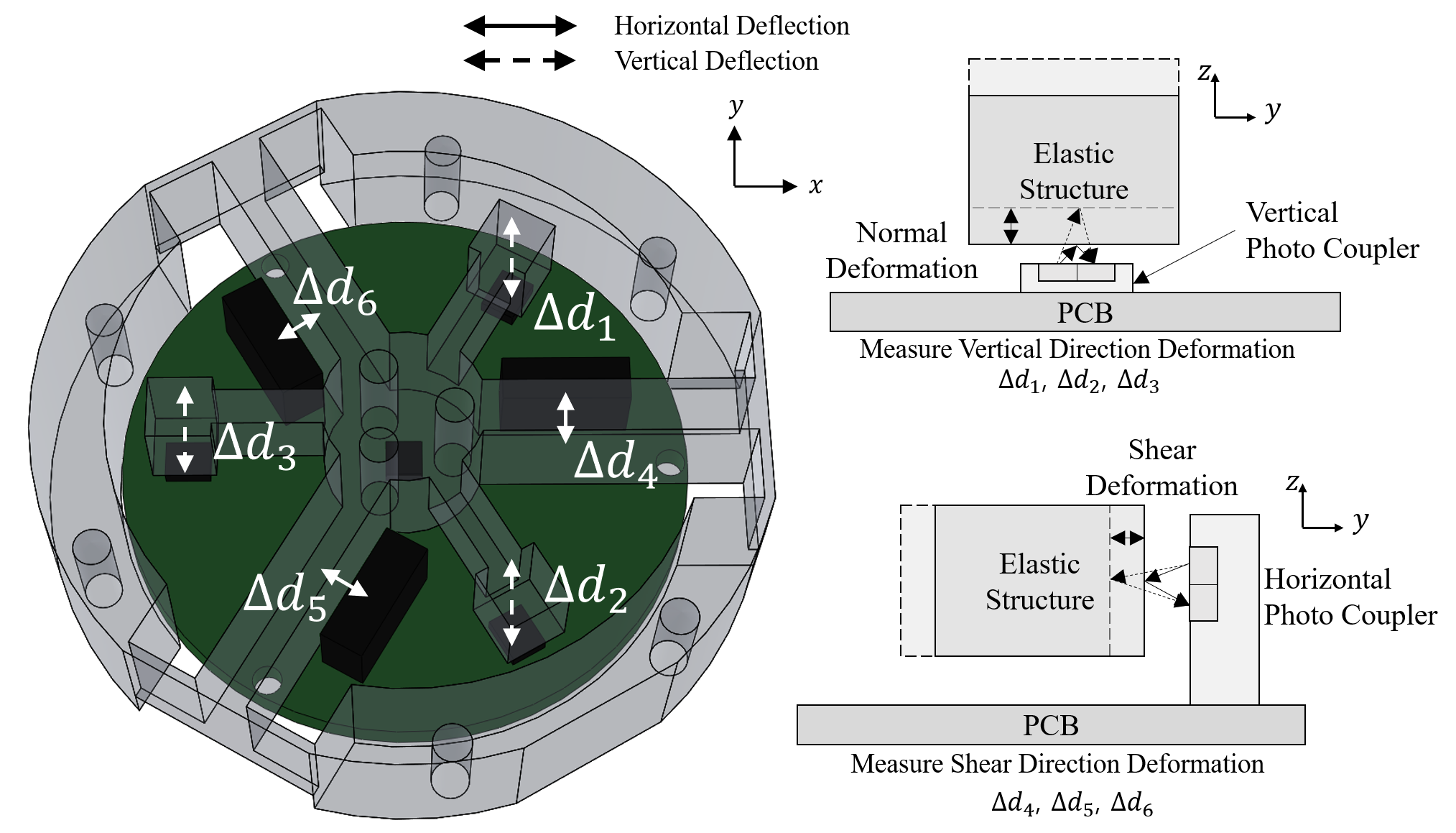}
    \caption{A novel six-axis force/torque sensor using photo-couplers: Distance between Vertical Photo-coupler and Reflective Surface of Elastic Structure $\Delta d_1,\Delta d_3,\Delta d_5$, Distance between horizontal Photo-coupler and Reflective Surface of Elastic Structure $\Delta d_2,\Delta d_4,\Delta d_6$ }
    \label{photocoupler}
\end{figure}
A novel six-axis force/torque sensor is presented in Fig.~\ref{photocoupler}. The measurement principle of this sensor is based on reflective photo-couplers. The sensor incorporates a total of six photo-couplers: three oriented vertically, primarily measuring the vertical force component (the force along the $z$-axis) and the moments about the $x$- and $y$-axes. The remaining three photo-couplers are oriented horizontally and are mainly used to measure the horizontal force components (the forces along the $x$- and $y$-axes) and the moment about the $z$-axis.

The sensor has a compact diameter of 40 mm and a lightweight design, weighing approximately 45 g. It integrates an Analog to Digital Converter(ADC), an Micro-Controller Unit(MCU), and a temperature sensor, enabling direct communication of the measured values via the CAN protocol. The measurement range is approximately 600 N for the forces along the $x$- and $y$-axes, 2000 N for the force along the $z$-axis, 14 N$\cdot$m for the moments about the $x$- and $y$-axes, and 20 N$\cdot$m for the moment about the $z$-axis. 

A TMP117 IC temperature sensor, which communicates via I2C, is mounted at the center of the printed circuit board. This sensor features an accuracy of 0.08°C and a resolution of 7.8125 m°C. In this study, it was utilized for temperature measurement and incorporated into the compensation algorithm.

Despite its small diameter and wide measurement range, the sensor achieves enhanced sensitivity through optimization via modeling and global search methods, providing a resolution of approximately 14 bits. The sensor is specifically designed for integration into the feet of legged robots, ensuring robustness against impact forces.

% a novel six-axis force/torque sensor는 Fig.~\ref{photocoupler}에 나와있다. 이 센서의 측정 원리는 반사형 포토커플러를 사용한 것이다. 총 6개의 포토 커플러가 있으며, 3개의 포토커플러는 수직 방향을 바라보고 있어 수직 힘의 종류인 z축힘과 x, y축 모멘트를 측정하는데 주로 사용된다. 나머지 3개는 수평 방향을 바라보고 있어 수평 힘의 종류인 x,y축 힘과 z축 모멘트를 측정하는데 주로 사용된다. 직경은 40mm이고 무게는 약 45g으로 가볍다. 이는 안에 ADC와 MCU, Temperature Sensor가 포함되어있어 바로 CAN통신을 이용하여 측정한 것을 바로 통신이 가능하다. 측정범위는 x,y축 힘이 약 600N이며, z축 힘은 2000N정도 되며, x,y축 모멘트는 14N.m정도 되고 z축 모멘트는 20Nm정도 된다. 여기서 측정 범위가 넓은데 직경이 작기 때문에 모델링과 global search방법을 이용하여 최적화를 진행하여 센서티비티를 증가하여 약 14bit 수준의 해상도를 가지고 있다. 이는 족형 로봇의 발에 장착하기 위해서 만들어져 임팩트에 강인한 성질을 가지고 있다.  

\section{Proposed Method and Learning of Network}
\begin{figure}[!thb]
    \centering
    \includegraphics[width=1\linewidth]{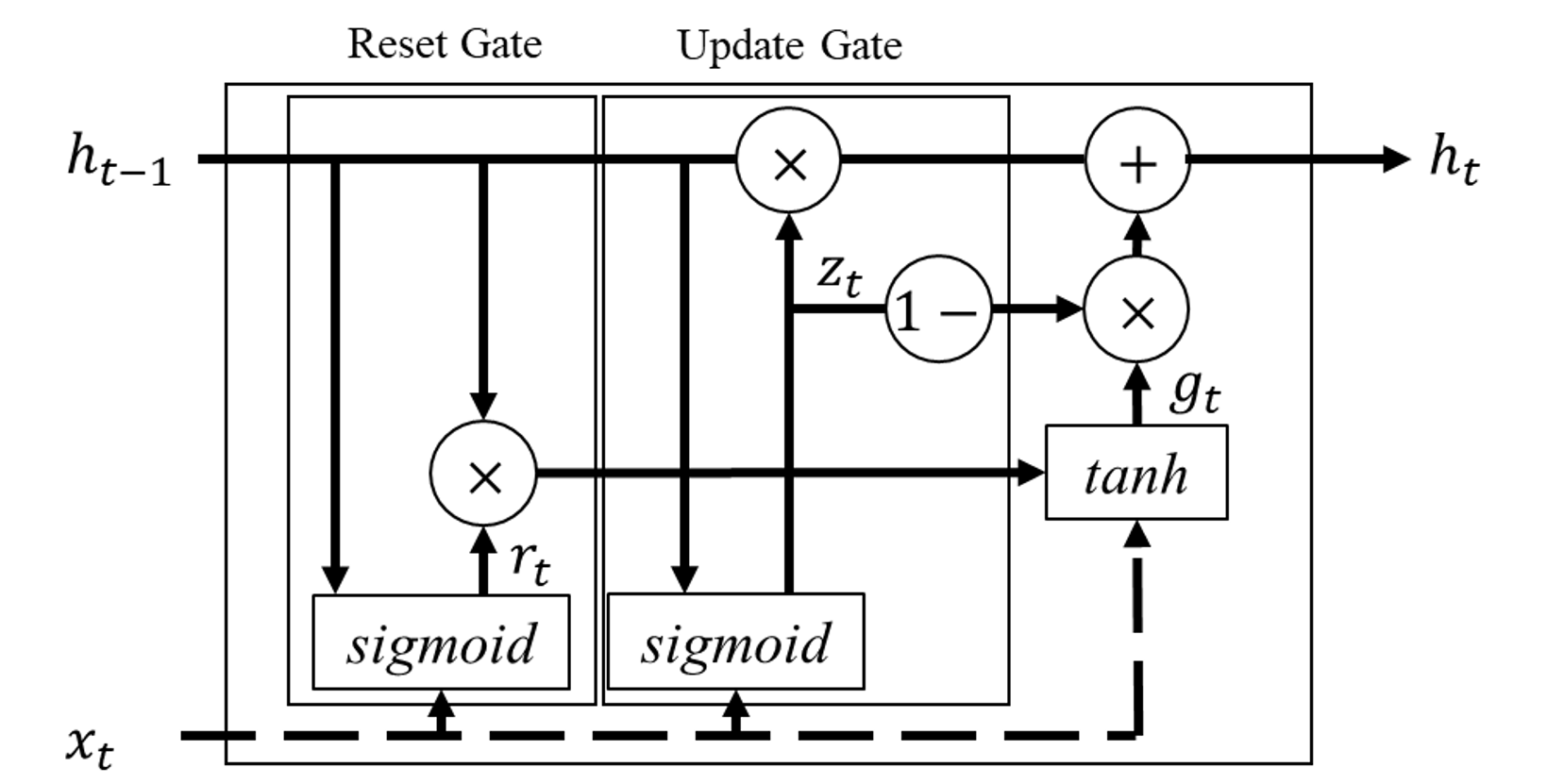}
    \caption{Structure of Gated Recurrent Unit}
    \label{mlptcngru}
\end{figure}
%  기존의 방법에서는 Multi-Layer Perceptron을 가장 많이 사용하였는데 이는 Fig.~\ref{mlptcngru}(a)에서 나온것 처럼 가장 기초적인 인공신경망으로 비선형성이 큰 모델을 학습시킬 때 좋은 network이다. 인풋과 아웃풋 layer가 있으며, 복수개의 Hidden layer가 있는 구조로 hidden layer에 활성화 함수를 ReLU나, Tanh 함수와 같은 비선형 함수를 사용해 파라미터를 학습시킨다. 
%  또한, Temporal Convolution Network는 Fig.~\ref{mlptcngru}(b)에서 나온 것 처럼 sequence input을 인풋으로 넣고 convolution을 이용하여 sequence data에서 특징을 잘 잡는 것으로 유명하며, 이는 짧은 시간의 데이터에서 특징을 잘 잡는다. 
% 제안하는 방법은 기존의 언어모델에서 많이 사용하는, RNN의 일종으로, LSTM과 비슷한 성능이지만 파라미터가 적은 GRU를 이용하여 온도 보상을 하는 방법이다. GRU는 Fig.~\ref{mlptcngru}(c)의 구조를 가지고 있으며, time sequence가 input $x_t$로 들어가서 이전의 output과 reset gate와 update gate를 이용하여 output을 만들어 내는 구조를 가지고 있다.
In conventional methods, the Multi-Layer Perceptron (MLP) has been the most widely used approach, as shown in Fig.~\ref{mlptcngru}(a). MLP is a fundamental artificial neural network architecture that is particularly effective for training models with high nonlinearity. It consists of an input layer, multiple hidden layers, and an output layer. The hidden layers utilize nonlinear activation functions such as ReLU or Tanh to optimize parameters.

Additionally, the Temporal Convolutional Network (TCN)~\cite{bai2018empirical}, as illustrated in Fig.~\ref{mlptcngru}(b), processes sequence inputs through convolutional operations, making it highly effective in capturing features from sequential data. TCN is known for its capability to extract meaningful features from short-term time series data.

The proposed method employs the Gated Recurrent Unit (GRU), a type of Recurrent Neural Network (RNN) commonly used in language models. GRU provides similar performance to Long Short-Term Memory (LSTM) networks~\cite{hochreiter1997long} but with fewer parameters, making it a more efficient choice for temperature compensation. The structure of GRU is depicted in Fig.~\ref{mlptcngru}(c), where the time sequence input \( x_t \) is processed using reset and update gates along with the previous output to generate the final output.

\begin{equation}
\begin{array}{l}
r_t = \sigma(W_{xr} x_t + W_{hr} h_{t-1}) \\
z_t = \sigma(W_{xz} x_t + W_{hz} h_{t-1}) \\
g_t = \tanh(W_{hg} (r_t \otimes h_{t-1}) + W_{xg} x_t) \\
h_t = (1 - z_t) \otimes g_t + z_t \otimes h_{t-1}
\end{array}
\label{GRUeq}
\end{equation}
% 리셋 게이트는 현재 상태에서 얼마나 이전 상태의 정보를 유지할지 결정하는 역할을 합니다. 값이 0에 가까울수록 이전 상태의 정보를 잊고, 1에 가까울수록 이전 상태의 정보를 기억한다. Fig.~\ref{mlptcngru}(c)의 update gate는 이전 상태의 정보와 새로운 정보를 가져오는 것 사이의 균형을 결정하는 역할을 한다. $z_t$는 0과 1사이의 실수 값으로 표현되며, 값이 1에 가까울수록 이전 상태의 정보를 우선적으로 가져오고, 0에 가까울 수록 새로운 정보를 우선적으로 가져온다. $z_t$가 1일 경우, 1 - $z_t$ 연산을 통과한 신호는 0이 되어 입력 게이트의 역할은 수행되지 않습니다. 값이 0일 경우, 망각 게이트는 수행되지 않고 입력 게이트만 수행된다. 이렇게 GRU는 이전상태의 정보와 현재 상태의 정보의 균형을 맞춰주는 역할을 하게 된다. Eq.~\ref{GRUeq}는 GRU 내부에서의 일어나는 게이트 연산을 수식으로 보여준다.
The reset gate determines how much of the previous state information should be retained in the current state. When the reset gate value is close to 0, the previous state information is forgotten, whereas when it is close to 1, the previous state information is preserved. The update gate in Fig.~\ref{mlptcngru}(c) plays a crucial role in balancing the contribution of previous state information and newly acquired information. 

The update gate value $z_t$ is a real number between 0 and 1. When $z_t$ is close to 1, the previous state information is prioritized, whereas when it is close to 0, the new information is emphasized. If $z_t$ is equal to 1, the signal passing through the operation $1 - z_t$ becomes zero, thereby disabling the input gate. Conversely, when $z_t$ is 0, the forget gate does not operate, and only the input gate is activated. 

Through this mechanism, the GRU effectively balances the previous and current state information. Eq.~\ref{GRUeq} presents the mathematical formulation of the gate operations within the GRU.

%또한, MLP를 이용하여 Time-Series Data를 직접 넣어주는 방식도 만들어서 비교를 하였다. MLP에서는 20개의 sequence Data를 넣어주었다.
% 기존의 MLP를 사용한 연구는 24nodes의 3개 레이어를 사용하는 것도 있으며, 36개의 node를 사용하는 것도 있다. 본 논문에서는 36개의 node와 3개의 layer를 이용하여 학습을 진행하였다. 여기서 Temporal Convolution Network도 이용하여 비교를 진행하였다. TCN은 time-series data의 특징을 잘 파악하고 robot의 상태 추정에도 사용이 많이 되고 있는 network 종류이다. 

\begin{figure}[!thb]
    \centering
    \includegraphics[width=1\linewidth]{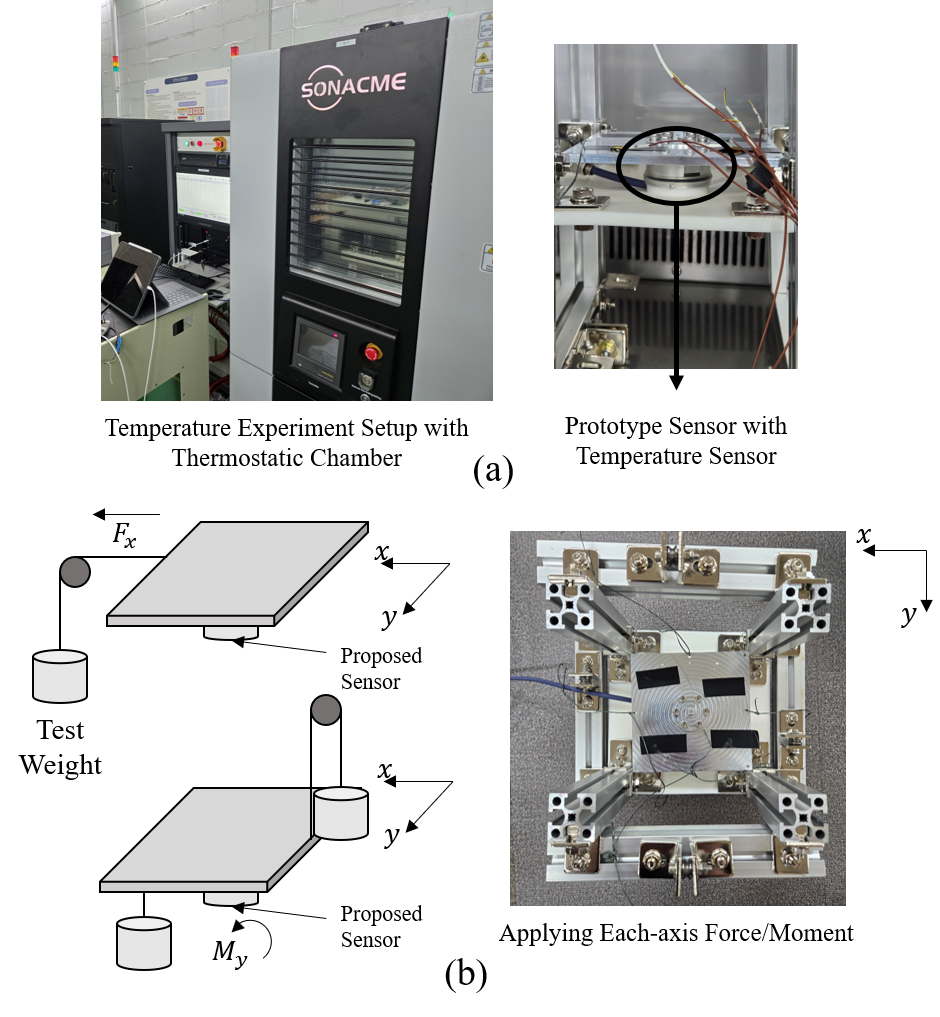}
    \caption{Temperature Data Acquisition Experiment:(a) Temperature Experiment Setup with Thermostatic Chamber (b) Equipment of Applying Static Force/Moment}
    \label{thermoexp}
\end{figure}

To evaluate the learning performance of conventional methods such as MLP, TCN, and GRU, an initial data acquisition experiment was conducted. The data was obtained using a thermostatic chamber, as illustrated in Fig.~\ref{thermoexp}. To assess the impact of forces along each axis on temperature compensation—specifically, the relationship between the sensor gain and temperature—the sensor was subjected to applied forces. Additionally, forces along each axis were measured using a test weight. The temperature was varied between -20°C and 60°C during the experiment.

\begin{figure}[!thb]
    \centering
    \includegraphics[width=1\linewidth]{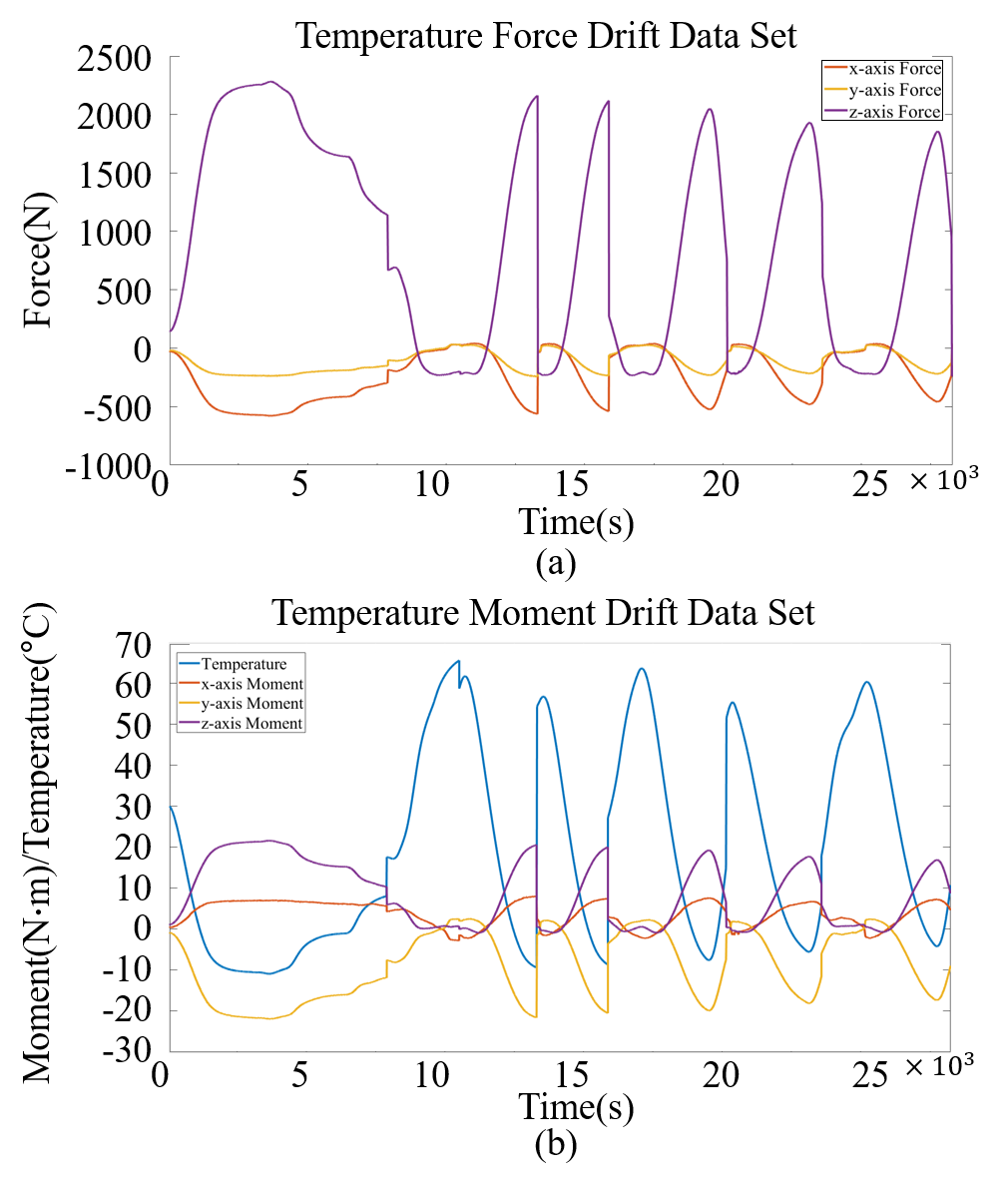}
    \caption{Temperature Drift Data Set: (a) Force (b) Moment}
    \label{tempdriftset}
\end{figure}

The data corresponding to temperature variations were measured by repeatedly cycling the temperature from -20°C to 60°C under normal conditions. The results are presented in Fig.\ref{tempdriftset}. As shown in Fig.\ref{tempdriftset}(b), the temperature was varied accordingly. The force drift for each output is depicted in Fig.\ref{tempdriftset}(a), while the moment drift is illustrated in Fig.\ref{tempdriftset}(b).
% 그렇게 Temperature 에 따른 데이터는 온도를 평상시에서 -20도에서 60도까지 여러번 반복을 하여 측정을 하였으며, 이는 Fig.~\ref{tempdriftset}에 나와있다. 온도는 Fig.~\ref{tempdriftset}(b)에서 나온 것처럼 변화를 시켰으며, 각 출의 Force drift는 Fig.~\ref{tempdriftset}(a)에서 나와있으며, Fig.~\ref{tempdriftset}(b)에서 Moment drift가 나와있다. 

\begin{figure}[!thb]
    \centering
    \includegraphics[width=1\linewidth]{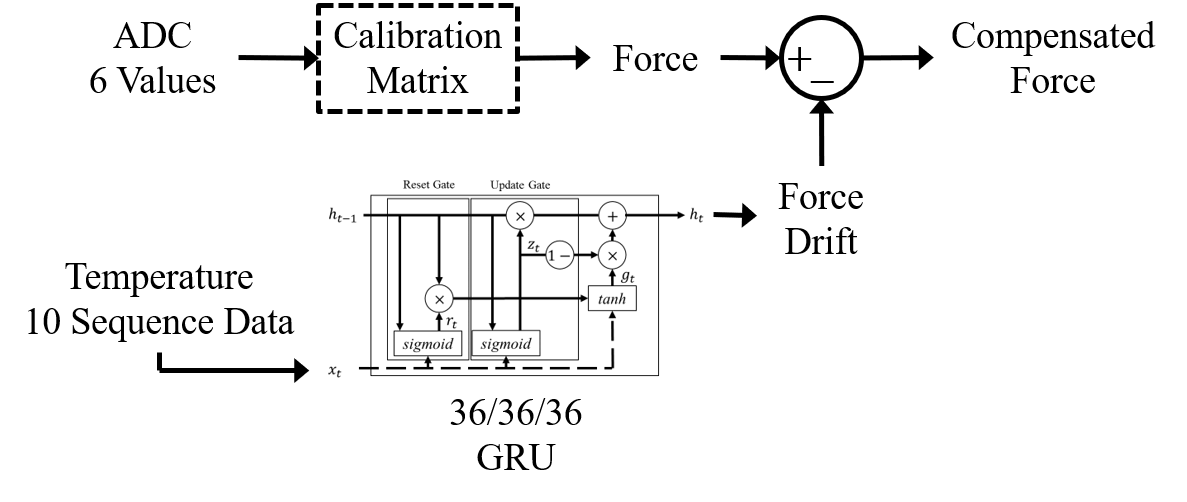}
    \caption{Diagram of Temperature Compensation Algorithm: Calculate Force Value from 6 ADC inputs and Calibration Matrix, Calculate Force Drift from 10 Sequence Temperature Sensor Data.  }
    \label{tempcomp}
\end{figure}
% Fig.~\ref{tempcomp}는 novel six axis force torque sensor에서 온도를 보상하는 방식에 대한 diagram을 보여준다. photo-couplers의 adc 값 6개를 사용하여 칼리브레이션 행렬을 거쳐 힘이 나오면 그것을 10개 sequence temperature data를 이용하여 학습된 learning 알고리즘을 이용하여 force/moment drift를 예측하고 그것을 이용하여 보상된 힘을 계산하는 형태로 이루어져 있다. Fig.~\ref{tempcomp}의 Gated Recurrent Unit 부분을 다른 네트워크인 MLP나 TCN을 이용해서 계산을 하면 다른 네트워크도 동일하게 적용할 수 있다. 
Fig.\ref{tempcomp} presents a diagram illustrating the temperature compensation method for the novel six-axis force/torque sensor. The process begins with the acquisition of six ADC values from the photo-couplers, which are then processed through a calibration matrix to compute the force. Subsequently, a learning algorithm, trained using a sequence of 10 temperature data points, predicts the force/moment drift. The compensated force is then calculated based on this prediction. Additionally, the Gated Recurrent Unit (GRU) in Fig.\ref{tempcomp} can be replaced with other network architectures, such as a Multi-Layer Perceptron (MLP) or a Temporal Convolutional Network (TCN), allowing for the application of different networks in the same framework.

\begin{figure*}[!thb]
    \centering
    \includegraphics[width=\linewidth]{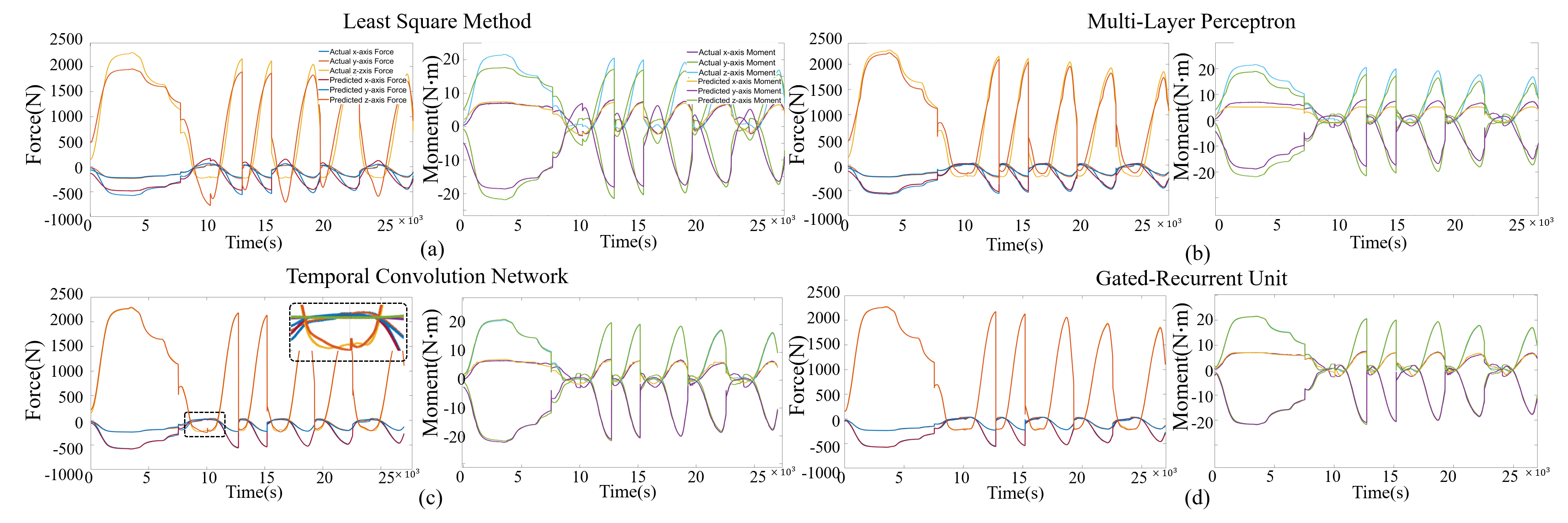}
    \caption{Comparison of learning performance across different methods (Least Square Method, MLP, TCN, GRU):
(a) Fitting results using the Least Square Method
(b) Fitting results using the Multi-Layer Perceptron (MLP)
(c) Fitting results using the Temporal Convolutional Network (TCN)
(d) Fitting results using the Gated Recurrent Unit (GRU)}
    \label{complsmmlp}
\end{figure*}

% 총 4가지의 방법을 학습을 비교를 하였는데 비교결과는 Fig.~\ref{complsmmlp}와 같다. Least Square Method와 36개 노드를 가지고 있는 3개 레이어를 가지고 있는 MLP와 TCN과 GRU를 각각 비교를 하였다. 
% Least Square Method는 비선형 성이 큰 온도 보상에서는 큰 오차를 보여주었으며, MLP도 온도 차이가 시간에 따라 발생함에 따라서 Fitting이 잘 되지 않는 것을 볼 수 있다. 그에 반해 Temporal Convolution Network나 Gated-Recurrent Unit을 보면 확실하게 더 잘 Fitting이 되는 것을 볼 수 있다. 하지만, TCN은 짧은 시간에서의 시간에 따른 데이터에 대해서 학습이 잘 되지만, 온도 변화는 오랜 시간이 걸리는 task이기 때문에 TCN보다 GRU가 더 확실하게 Fitting이 잘 되는 것을 보였다. 
A comparison of the four learning methods is presented in Fig.~\ref{complsmmlp}. The Least Square Method, a Multi-Layer Perceptron (MLP) with three layers and 36 nodes per layer, a Temporal Convolutional Network (TCN), and a Gated Recurrent Unit (GRU) were evaluated. In this study, the input consists of temperature data, including both single data points and sequences, while the output represents the estimated force drift. The Adam optimizer was used for training, with a maximum of 2000 epochs. The learning rate was set to 0.001, the batch size was 128, and the mean squared error (MSE) was employed as the loss function.

The Least Square Method exhibited significant errors in temperature compensation due to its inability to handle strong nonlinearity. Similarly, the MLP struggled to achieve proper fitting as temperature variations occurred over time. In contrast, both the TCN and GRU demonstrated significantly better fitting performance. However, while the TCN effectively captured short-term temporal dependencies, temperature variations occur over extended periods, making the GRU more suitable for this task. As a result, the GRU achieved the most accurate fitting among the evaluated methods.

% 여기서 input은 Temperature(Single Data and Sequence) 이며, output은 estimated force drift이고 ADAM을 사용하였으며, Max Epoch는 2000으로 정했고 Learning rate는 0.001이며, batch size는 128이고 Loss Function은 MSE를 사용하였다.

\begin{table}[!htb]
    \centering
    \caption{RMSE After Convergence for Each Method}
    \begin{tabular}{ccccc}
    \hline
    \hline
    Method     &LSM&MLP&TCN&GRU  \\ \hline
    RMSE     &$\sim$500&$\sim$100&$\sim$40&$\sim$5
    \\\hline
    \end{tabular}
    \label{tab1}
\end{table}

% Table~\ref{tab1}을 보면 LSM은 RMSE가 500가까이 되어서 오차가 너무 크며, MLP도 100 가까이 나오는 것을 알 수 있다. TCN이 40이고 GRU는 5미만으로 LSM 기준으로 100분의 1 수준의 크기의 RMSE로 GRU가 가장 수렴이 잘되고 Fitting이 잘 되는 것을 확인 할 수 있다.
As shown in Table~\ref{tab1}, the RMSE for the Least Square Method (LSM) is close to 500, indicating a large error. The MLP also exhibits a relatively high RMSE of approximately 100. In contrast, the TCN achieves an RMSE of 40, while the GRU achieves an RMSE below 5. This demonstrates that the GRU achieves the best convergence and fitting performance, with an RMSE that is nearly 1/100th of that of the LSM.

\section{Algorithm Validation Experiment}

\begin{figure}[!thb]
    \centering
    \includegraphics[width=1\linewidth]{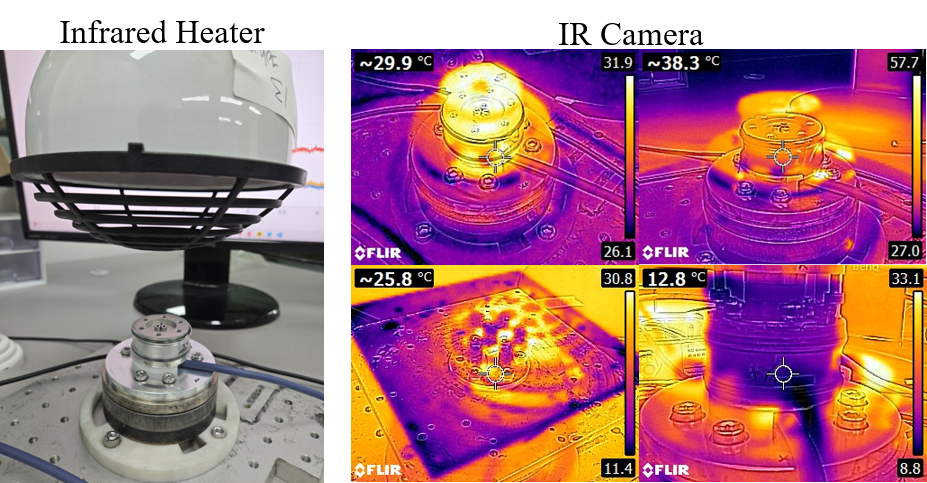}
    \caption{Temperature Compensation Experiment using Infrared Heater and Ice for Verification }
    \label{testver}
\end{figure}

% Fitting 한 함수가 기존의 thermostatic chamber에서 얻은 환경이 아닌 실제 환경에서 작동하는지 확인하기 위해서 실험을 진행하였다. 실험은 이전 연구에서 활용했던 방법인 Infrared Heater를 사용하였으며, 온도를 낮추기 위해서는 얼음주머니를 사용하였다. 실험은 두가지 방식으로 이루어졌다. 먼저 온도를 높이는 실험, 그리고 온도를 낮추는 실험을 진행하였다. 온도를 올리는 실험은 약 200초 동안 10도를 올리는 실험을 진행하였으며, 온도를 낮추는 실험은 약 300초 동안 5도를 낮추는 속도로 진행되었다. 
% 실험 결과는 여기서 Fig.~\ref{tempexp}에 나와있다. 여기서 비교를 한 것은 Compensation을 하지 않은 것과 MLP를 사용한 것, 그리고 10개 시간 순서 온도 데이터를 이용하여 MLP를 학습한 것과 Gated-Recurrent Unit을 사용한 것을 비교를 하였다. 여기서 10개 시간 순서 온도 데이터를 이용한 MLP는 TCN과 비슷하게 RMSE가 40정도로 기존의 MLP보다 더 학습이 잘 되었다.

An experiment was conducted to verify whether the fitted function operates effectively in real-world conditions rather than in the controlled environment of a thermostatic chamber. The experiment utilized an infrared heater, as employed in previous study~\cite{billeschou2021low}, to increase the temperature, while an ice pack was used to decrease it.

The experiment consisted of two parts: one for increasing the temperature and another for decreasing it. In the heating experiment, the temperature was increased by 10°C over approximately 200 seconds, while in the cooling experiment, the temperature was decreased at a rate of 5°C over approximately 300 seconds.

\begin{figure*}[!thb]
    \centering
    \includegraphics[width=1\linewidth]{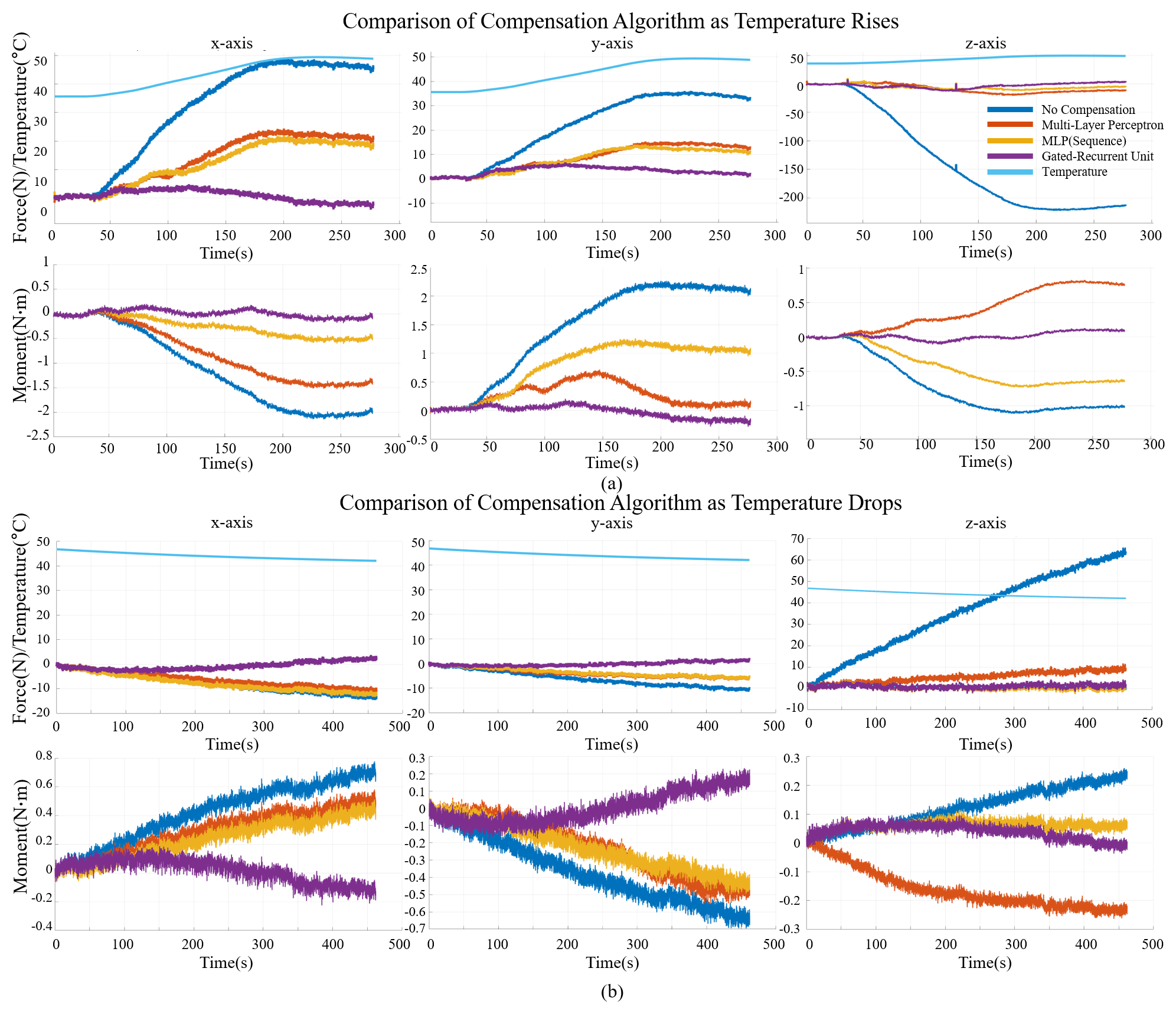}
    \caption{Comparison of Compensation Algorithm as Temperature Changes: (a) Experiment Results at Temperature Rises (b) Experiment Results at Temperature Drops}
    \label{tempexp}
\end{figure*}

The experimental results are presented in Fig.~\ref{tempexp}. The comparison includes cases without compensation, with compensation using an MLP, with an MLP trained using a sequence of 10 temperature data points, and with a Gated Recurrent Unit (GRU). The MLP trained with 10 sequential temperature data points showed improved learning performance, achieving an RMSE of approximately 40, similar to the TCN, and outperforming the conventional MLP.

Fig.\ref{tempexp}(a) presents the experimental results when using the temperature compensation method described in Fig.\ref{tempcomp} during temperature increase, with no external force applied to any axis. The results show that the compensation method using the GRU effectively eliminates drift across all axes.

In the z-axis force, both the GRU and the MLP trained with sequential temperature data exhibit similar performance. However, the MLP demonstrates a tendency to learn nonlinearities better than the GRU. This is likely due to the fact that the vertical photo-coupler, responsible for the z-axis measurement in the novel force/torque sensor, is positioned farther from the reflective surface compared to the horizontal photo-couplers. Consequently, it is more susceptible to ambient dark current effects, resulting in greater nonlinearity.

Fig.~\ref{tempexp}(b) presents the experimental results for temperature compensation during temperature decrease, again with no external force applied to any axis. The results confirm that the GRU exhibits the least drift, providing significantly better compensation compared to the MLP.
% Fig.~\ref{tempexp}(a)는 온도가 올라갈 때 Fig.~\ref{tempcomp}와 같은 온도 보상 방식을 사용할 때의 실험 결과를 보여주며 각 축에 힘을 주지 않았을 때 데이터를 보여준다. 여기서 모든 축에서 GRU를 이용한 보상 방식은 Drift가 거의 없는 보습을 보여준다. 이때 z축 힘에서는 GRU와 Sequence를 사용한 MLP가 비슷한 영향을 보여주었는데 GRU보다 MLP가 비선형성에 대해서 더 잘 학습되는 경향이 있으며, novel force/torque sensor의 z축을 담당하는 vertical photo-coupler가 horizontal photo-coupler보다 reflective surface에 멀리 위치해 있기 때문에 주위의 dark current의 영향을 더 받기 때문에 비선형성이 크다. Fig.~\ref{tempexp}(b)는 온도가 내려갈 때의 온도 보상 실험 결과를 보여주며, 각 축에 힘을 주지 않았을 때 데이터를 보여준다. 여기서도 GRU가 Drift가 제일 덜 생겼으며, 확실하게 MLP를 사용한 것 보다 더 보상이 잘 되는 것을 확인 할 수 있다.  

\begin{table}[!h]
\centering
\caption{RMS Error Comparison between Compensation Method when Temperature Decreases}
\resizebox{0.5\textwidth}{!}{
\begin{tabular}{ccccccc}
\hline\hline
RMSE       & $F_x$(N)      & $F_y$(N)      & $F_z$(N)      &$M_x$(N$\cdot$m)         & $M_y$(N$\cdot$m)        & $M_z$(N$\cdot$m)        \\ \hline
No compensation           & 8.6585   & 6.8711   & 40.009   & 0.4673      & 0.4128      & 0.1409     \\ 
MLP      & 6.8087   & 4.2006   & 5.8157   & 0.3293       & 0.2819       & 0.1750      \\ 
MLP(Sequence)      & 8.3733  & 3.9628   & 0.9649   & 0.2717       & 0.2636       & 0.0631      \\ 
GRU & 1.7778 & 0.8667 & 1.2555& 0.0824 & 0.0903 & 0.0454 \\ \hline
\end{tabular}}

\label{RMSED}
\end{table}
\begin{table}[!h]
\centering
\caption{RMS Error Comparison between Compensation Method when Temperature Increases}
\resizebox{0.5\textwidth}{!}{
\begin{tabular}{ccccccc}
\hline\hline
RMSE       & $F_x$(N)      & $F_y$(N)      & $F_z$(N)      &$M_x$(N$\cdot$m)         & $M_y$(N$\cdot$m)        & $M_z$(N$\cdot$m)        \\ \hline
No compensation           & 35.381   & 25.317   & 159.82   & 1.3810      & 1.6356      & 0.8197     \\ 
MLP      & 16.003   & 10.039   & 11.315   & 0.9624       & 0.3397       & 0.4881      \\ 
MLP(Sequence)      & 14.086  & 9.1935   & 7.0742   & 0.3259       & 0.8862       & 0.5026      \\ 
GRU & 2.1518 & 3.7264 & 5.1868& 0.0749 & 0.1031 & 0.0557 \\ \hline
\end{tabular}}

\label{RMSEI}
\end{table}

% Table.~\ref{RMSED}는 Fig.~\ref{tempexp}(a)의 결과를 RMSE를 표로 나타낸 것이다. 여기서 GRU는 모든 축에서 균일하게 좋은 성능을 보여주었으며, z축 힘에서 Sequence를 사용한 MLP보다 약 0.3N정도 차이가 나는 것을 볼 수 있지만 전체적인 성능은 약 4배 좋은 것을 볼 수 있다. Table.~\ref{RMSEI}는 Fig.~\ref{tempexp}(b)의 결과를 RMSE를 표로 나타낸 것이며, 여기서는 GRU가 모든 축에서 더 좋은 결과를 보여주었다. 특히 z축 힘과 같은 경우에 compensation을 하지 않았을 때 160N이 나오지만 약 5N의 RMSE로 줄어든 것을 볼 수 있으며, x축 힘에서도 35N의 drift에서 2N수준의 drift로 감소한 것을 확인 할 수 있다. 이는 약 최대 0.2\% Full-scale 오차를 보여준다.  
Table~\ref{RMSED} presents the RMSE values corresponding to the results in Fig.~\ref{tempexp}(a). The GRU demonstrated consistently superior performance across all axes. Although there was a slight difference of approximately 0.3 N in the z-axis force compared to the MLP trained with sequential data, the overall performance of the GRU was approximately four times better.

Table~\ref{RMSEI} summarizes the RMSE values for the results shown in Fig.~\ref{tempexp}(b). In this case, the GRU achieved the best performance across all axes. Notably, in the z-axis force, the RMSE was reduced from 160 N without compensation to approximately 5 N. Similarly, in the x-axis force, the drift was reduced from 35 N to approximately 2 N. This corresponds to a maximum full-scale error of approximately 0.2\%.

\begin{figure*}[!thb]
    \centering
    \includegraphics[width=1\linewidth]{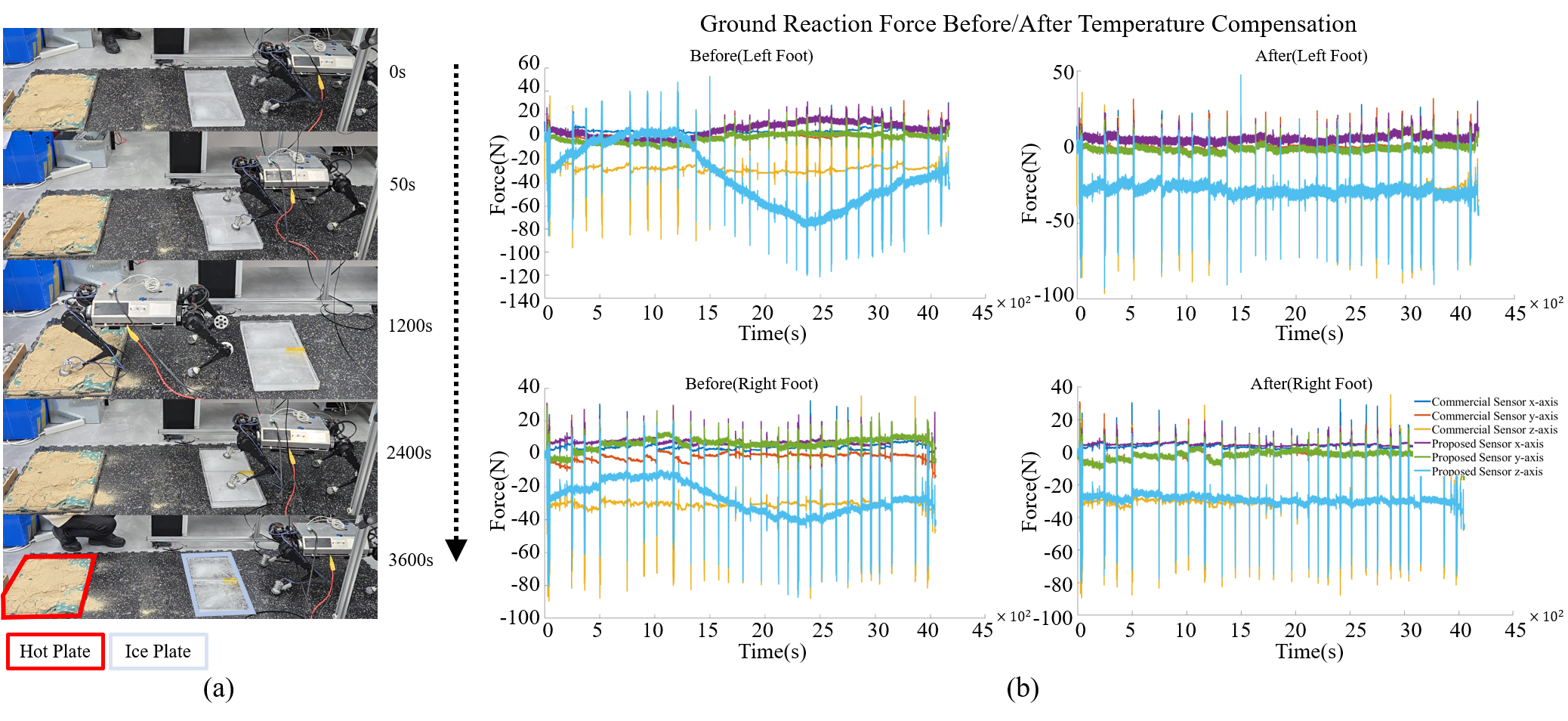}
    \caption{Temperature Compensation Experiment in Quadruped Application: (a) Walking Experiment Ice Plate and Hot Plate (b) Graph of Ground Reaction Force Before/ After Temperature Compensation}
    \label{quadexp}
\end{figure*}

In the previous experiments, tests were conducted using the sensor alone. However, to further validate its real-world applicability through comparative analysis, an experiment was performed by mounting both a commercial sensor and the novel force/torque sensor on the foot of a quadruped robot using GRU-based temperature compensation algorithm. This setup allowed for direct comparison and evaluation in practical conditions.

For the commercial sensor, the RFT-40 model from Robotus was used. This capacitive-type sensor was selected due to its minimal sensitivity to temperature variations. The experiment setup is illustrated in Fig.~\ref{quadexp}.

Fig.~\ref{quadexp}(a) depicts the experimental procedure, where the robot initially stepped onto a -20°C ice plate. After 20 minutes, it moved onto a 70°C hot plate, heated using hand warmers, and remained there for another 20 minutes. The robot then stepped back onto a new ice plate at -20°C for an additional 20 minutes, replicating realistic temperature variations in a controlled manner.

Fig.~\ref{quadexp}(b) presents the measured ground reaction forces (GRFs) of the left and right feet during the experiment. The graph on the left shows the GRF before applying temperature compensation, while the graph on the right illustrates the results after compensation.

% 앞에서 센서 혼자서는 실험을 진행하였지만, 다른 센서와 비교를 하면서 실제 적용하는 실험을 통해 검증을 더 확실하게 하기 위해 사족 로봇의 발에 commercial sensor와 novel force/torque sensor를 같이 달아서 실험을 진행하였다. 여기서 Commercial Sensor는 Robotus의 RFT-40 모델을 사용하였으며, Capacitive type으로 온도영향을 거의 받지 않는 센서를 사용하였다. 그렇게 실험을 Fig.~\ref{quadexp}와 같이 하였으며, Fig.~\ref{quadexp}(a)는 로봇이 처음에 -20도의 iceplate를 밟고 있다가 20분 뒤에 핫팩으로 만든 70도의 hot plate위에 20분 있다가 뒤로 다시 가서 새로운 iceplate를 20분동안 밟아 온도 변화를 실제 환경과 비슷하게 재현하여 실험을 진행하였다. 이때, Fig.~\ref{quadexp}(b)는 왼발과 오른발의 Ground Reaction Force를 실험에서 측정한 것을 보여주는 그래프로 왼쪽의 온도 보상하기전에서 보상을 한 후의 그래프를 보여준다. 

%여기서 기존의 MLP, TCN, GRU 방법이 얼마나 학습이 잘되는지 확인하기 위해 먼저 데이터 취득실험을 하였다. 데이터는 Fig.~\ref{thermoexp}와 같이 thermostatic chamber에서 취득을 하였다. 여기서 각 축의 힘이 온도 보상에 영향을 주는지 확인하기 위해서 즉 센서의 gain과 온도간의 영향을 확인하기 위해서 센서에 힘을 줬을 때와 각 축의 힘을 test weight를 이용하여 힘을 줬을 때를 측정을 하였다. 여기서 온도는 섭씨 -20도에서 60도로 설정하여 실험을 진행하였다.

\begin{table}[!h]
\centering
\caption{RMS Error Comparison between Before and After Compensation during Walking Experiment}
\begin{tabular}{cccc}
\hline\hline
RMSE        & $F_x$(N)      & $F_y$(N)      & $F_z$(N)     \\ \hline
  Left Foot(Before)         & 6.280   & 3.084   & 25.502    \\ 
Left Foot(After) & 1.224      & 2.145 & 2.714\\ 
Right Foot(Before)       & 3.399 & 8.274 & 9.864      \\ 
  Right Foot(After)       &   1.369 & 1.136 & 2.831 \\ \hline
\end{tabular}
\label{RMSEcompensation}
\end{table}

As a result of the experiment, the RMSE values were calculated as shown in Table~\ref{RMSEcompensation}. For the sensor on the left foot, the RMSE of the z-axis force was 26 N without temperature compensation. However, after applying compensation, it was reduced to 2.7 N, nearly a tenfold improvement. A similar trend was observed for the right foot, confirming that the temperature compensation method was effectively applied.

% 실험을 한 결과 Table.~\ref{RMSEcompensation}와 같이 RMSE를 계산할 수 있었다. 왼쪽 발의 센서에서는 온도 보상을 하지 않았을 때 z축 힘에서 26N의 RMSE가 나왔지만, 온도 보상을 한 후에는 2.7N으로 거의 10배 가까이 적어진 것을 볼 수 있다. 오른 쪽 발도 비슷한 결과를 얻을 수 있어 온도보상이 확실하게 적용이 되는 것을 확인 할 수 있었다.  

\section{Conclusion}

This study proposed a temperature compensation algorithm for a novel six-axis force/torque sensor using a Gated Recurrent Unit (GRU)-based learning approach. Unlike conventional methods such as the Least Square Method (LSM) and Multi-Layer Perceptron (MLP), which struggle with nonlinear temperature-induced drift, the GRU effectively captured temporal dependencies, enabling robust compensation even under continuous temperature variations.

To validate the proposed method, multiple experiments were conducted. Initially, controlled experiments in a thermostatic chamber demonstrated that the GRU significantly outperformed other approaches, achieving an RMSE nearly 100 times lower than LSM and approximately four times lower than MLP. Further validation was performed in real-world conditions using an infrared heater and an ice pack, confirming the effectiveness of the GRU-based compensation compared to MLP and Temporal Convolutional Networks (TCN). Additionally, an application experiment was conducted by integrating the sensor into the foot of a quadruped robot alongside a commercial capacitive force sensor. The results showed that temperature drift, which initially reached 26 N in the z-axis, was reduced to 2.7 N after compensation, demonstrating nearly a tenfold improvement.

These findings indicate that the proposed GRU-based temperature compensation method effectively mitigates the impact of temperature-induced drift, even in small and lightweight force/torque sensors that are highly susceptible to thermal fluctuations. The proposed approach has the potential to be applied to various sensors requiring robust compensation against nonlinear temperature effects, making it particularly beneficial for applications in legged robots and other robotic systems operating in dynamic environments. Future research could explore further optimization of network architectures and extend the method to compensate for additional environmental factors affecting sensor performance.

% \section*{Appendix}

% Appendixes, if needed, appear before the acknowledgment.

% \section*{Acknowledgment}

% The preferred spelling of the word ``acknowledgment'' in American English is without an ``e'' after the ``g.'' Use the singular heading even if you have many acknowledgments. Avoid expressions such as ``One of us (S.B.A.) would like to thank ... .'' Instead, write ``F. A. Author thanks ... .'' In most cases, sponsor and financial support acknowledgments are placed in the unnumbered footnote on the first page, not here.

% References

\bibliographystyle{Bibliography/IEEEtranTIE}

% \vspace{-1cm}
\begin{IEEEbiography}[{\includegraphics[width=1in,height=1.25in,clip,keepaspectratio]{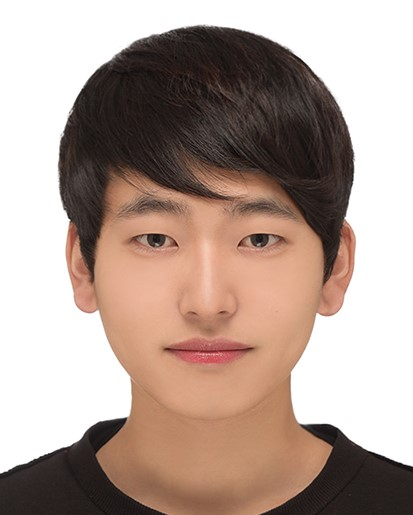}}]
{Hyun-Bin Kim}
~ received the B.S., M.S. and Ph.D. degrees in mechanical engineering from Korea Advanced Institute of Science and Technology(KAIST), Daejeon, Republic of Korea, in 2020, 2022 and 2025 respectively. He is currently working as the post-doctor researcher in KAIST. His current research interests include force/torque sensors, legged robot control, robot design and mechatronics system.
\end{IEEEbiography}
\begin{IEEEbiography}[{\includegraphics[width=1in,height=1.25in,clip,keepaspectratio]{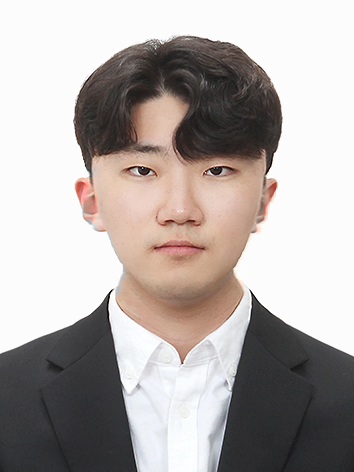}}]
{Seokju Lee}~ received the B.S. degree in electrical engineering from Ulsan National Institute of Science and Technology (UNIST), Ulsan, Republic of Korea, in 2023, and the M.S. degree in mechanical engineering from Korea Advanced Institute of Science and Technology (KAIST), Daejeon, Republic of Korea, in 2025. He is currently working toward the Ph.D. degree with the Department of Mechanical Engineering, KAIST.
His research interests include legged robots, state estimation, and reinforcement learning.
\end{IEEEbiography}
\begin{IEEEbiography}[{\includegraphics[width=1in,height=1.25in,clip,keepaspectratio]{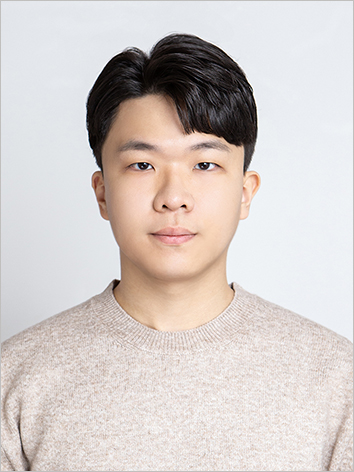}}]{Byeong-Il Ham}
~ received the B.S. degree in school of robotics from University of Kwangwoon, Seoul, and the M.S. degree in robotics program from Korea Advanced Institute of Science and Technology(KAIST), Daejeon, Republic of Korea, in 2022 and 2024, respectively. He is in Doctor Program in KAIST, Daejeon, Korea, from 2024. His current research interests include legged system, optimal control and motion planning.
\end{IEEEbiography}
% \vspace{11pt}
% \begin{IEEEbiography}[{\includegraphics[width=1in,height=1.25in,clip,keepaspectratio]{CKH_picture.jpg}}]{Keun-Ha Choi}
% ~ received the B.S. degree in weapon system engineering from Korea Military Academy, Seoul, and the M.S. degree and the Ph.D. degrees in mechanical engineering from Korea Advanced Institute of Science and Technology(KAIST), Daejeon, Republic of Korea in 2002, 2007 and 2016 respectively. He was a Defense Acquisition Program Administration, a Project Management Officer, Republic of Korea, from 2009 to 2012 and from 2016 to 2019, Army Education and Doctrine Command, an AI Research Officer, from 2019 to 2020, and an Army Headquarters, Force Planning Officer, from 2020 to 2021. In 2021, he was an AI/Big Data Research Officer with the Army Future Innovation Research Center. In 2022, he joined as a Research Assistant Professor with Daedong-KAIST Research Center for Mobility. Since 2023, he has been with the Department of Mechanical Engineering, KAIST. His current research interests include sensor fusion-based robot autonomous navigation algorithm and control, vision sensor-based object detection using AI, and defense AI application plan/military operation concept/concept design.
% \end{IEEEbiography}
\vspace{11pt}
\begin{IEEEbiography}[{\includegraphics[width=1in,height=1.25in,clip,keepaspectratio]{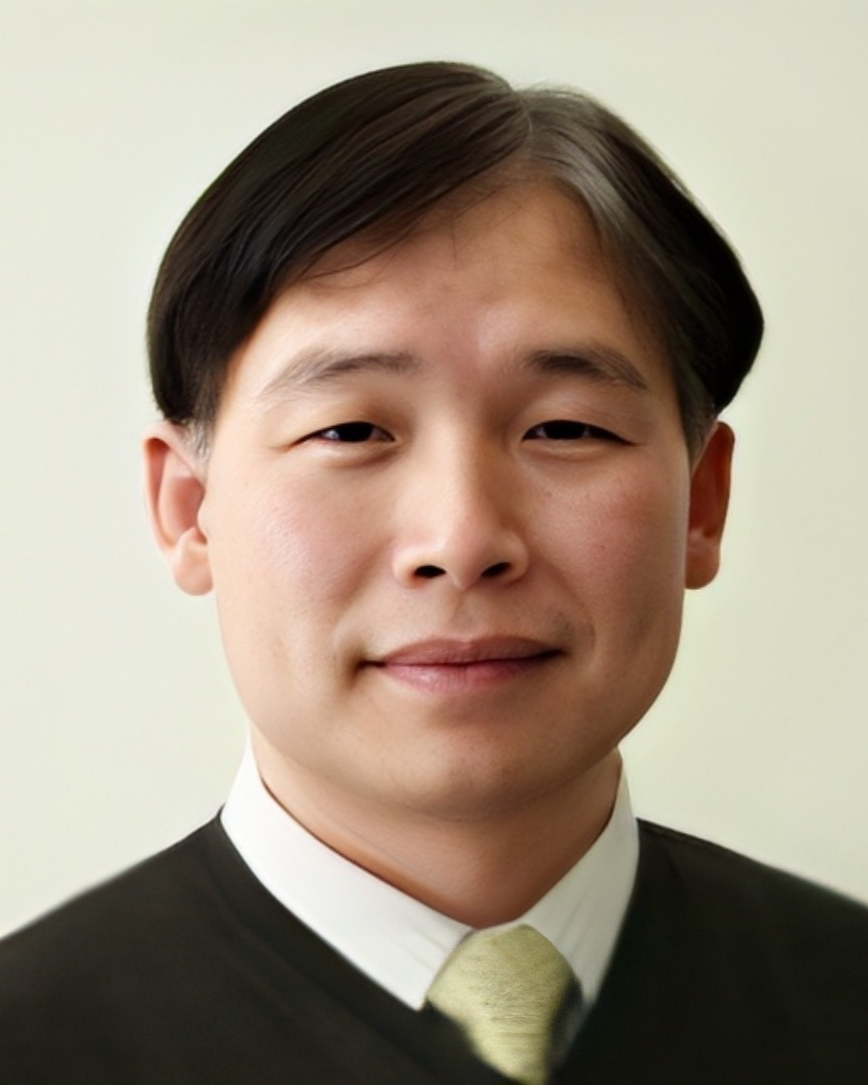}}]{Kyung-Soo Kim}~(Fellow, IEEE)~ received the B.S., M.S., and Ph.D. degrees in mechanical engineering from Korea Advanced Institute of Science and Technology (KAIST), Daejeon, Republic of Korea, in 1993, 1995, and 1999, respectively. He was a Chief Researcher with LG Electronics Inc., from 1999 to 2003, and the DVD Group Manager of STMicroelectronics Company Ltd., from 2003 to 2005. In 2005, he joined the Department of Mechanical Engineering, Korea Polytechnic University, Siheung, Republic of Korea, as a Faculty Member. Since 2007, he has been with the Department of Mechanical Engineering, KAIST. His research interests include control theory, electric vehicles, and autonomous vehicles. He serves as an Associate Editor for the Automatica and the Journal of Mechanical Science and Technology.
\end{IEEEbiography}

\end{document}